\documentclass[runningheads]{llncs}

\usepackage[T1]{fontenc}
\def\doi#1{\href{https://doi.org/\detokenize{#1}}{\url{https://doi.org/\detokenize{#1}}}}
\usepackage{graphicx}
\usepackage{listings}
\usepackage{amsmath,amssymb,amsfonts}
\usepackage{algorithmic}
\usepackage{graphicx}
\graphicspath{ {./figures/} }
\usepackage{textcomp}
\DeclareMathOperator*{\argmax}{arg\,max}

\usepackage{url}

\usepackage[dvipsnames]{xcolor}

\begin{document}

\title{A deep learning network with differentiable dynamic programming for retina OCT surface segmentation}
\titlerunning{A DL network with DDP for OCT segmentation}
%
\author{Hui Xie \and
Weiyu Xu \and
Xiaodong Wu}
\authorrunning{H. Xie et al.}
%
\institute{The University of Iowa, Iowa City, IA 52242, USA\\
\email{xiaodong-wu@uiowa.edu}
}
\maketitle              

\begin{abstract}
Multiple-surface segmentation in Optical Coherence Tomography (OCT) images is a challenge problem, further complicated by the frequent presence of weak image boundaries. Recently, many deep learning (DL) based methods have been developed for this task and yield remarkable performance. Unfortunately, due to the scarcity of training data in medical imaging, it is challenging for DL networks to learn the global structure of the target surfaces, including surface smoothness. To bridge this gap,  this study proposes to seamlessly unify a U-Net for feature learning with a constrained differentiable dynamic programming module to achieve an end-to-end learning for retina OCT surface segmentation to explicitly enforce surface smoothness.
It effectively utilizes the feedback from the downstream model optimization module to guide feature learning, yielding a better enforcement of global structures of the target surfaces. Experiments on Duke AMD (age-related macular degeneration) and JHU MS (multiple sclerosis) OCT datasets for retinal layer segmentation  demonstrated very promising segmentation accuracy.

\end{abstract}

\section{Introduction} \label{sec:introduction}
\vspace{-4mm}
Highly accurate surface segmentation for retina optical coherence tomography (OCT) is a clinical necessity in many diagnostic and treatment tasks of ophthalmic diseases. In retina OCT imaging, the frequent presence of weak image boundaries complicated by image artifacts often leads to undesirable boundary spikes with many automated segmentation methods. However, experienced ophthalmologists can well delineate retinal surfaces from OCT scans in those difficult scenarios  while taking advantage of their global shape information and mutual interaction. This indicates that surface insufficiency can be remedied by  making use of surface shape and context priors in the segmentation methods~\cite{Songqi_GraphSearch_2012}. We thus propose to seamlessly integrate differentiable dynamic programming (DDP)~\cite{Mensch_DDP_2018} into a deep learning framework with an end-to-end training for retina OCT surface segmentation to enforce surface smoothness.

Many retina OCT segmentation methods have been proposed in past years. Garvin {\em et al.} first introduced the graph-based optimal surface segmentation method~\cite{Likang_GraphSearch_2005} for surface delineation in retinal OCT~\cite{Garvin_MacularOCTSeg_2008}, which was further developed by incorporating various {\em a priori} knowledge reflecting anatomic and imaging information~\cite{Songqi_GraphSearch_2012,Shah_ConvexPrior_2019}. Other known OCT surface segmentation approaches include level set~\cite{carass2014multiple,gawlik2018active,liu2019layer}, probabilistic global shape model~\cite{rathke2014probabilistic}, random forest classifier~\cite{Lang_AURA_2013,xiang2018automatic}, and dynamic programming~\cite{chiu2010automatic,yang2010automated,Keller_ShortPathGraphSearch_2016,rathke2017_DPProbabilisticModel}. Each of these traditional methods has its own strength. They all share a common drawback that is their dependence on handcrafted features.


Armed with superior data representation learning capacity, deep learning (DL) methods are emerging as powerful alternatives to traditional segmentation algorithms for many medical image segmentation tasks~\cite{litjens2017survey,shen2017deep}. Fully convolutional networks (FCNs)~\cite{schlegl_FCNMacularFluidSeg_2018,masood2019automatic}, Convolutional neural networks (CNNs)~\cite{Abhay_CNN_Surface_2018},  and U-Net~\cite{Roy_RelayNet_2017,lee2017deep,YufanHe_MIA_2021,LiuHong_HybridNetworkOCT_2021,Xie_IPMSurface_2022} have been utilized for retinal layer segmentation in OCT images.  Due to the scarcity of training data in medical imaging, it is yet nontrivial for DL networks to {\em implicitly} learn  global structures of the target surfaces. Thus, the retinal layer topology cannot be guaranteed with those methods, neither the continuity and smoothness of the retinal surfaces can be ensured. To address those limitations, the graph-based method and dynamic programming were used as post-processing for the deep learning models to enforce surface monotonicity and smoothness~\cite{Fang_GraphSearchDL_2017,Kugelman_RNNGraphSearch_2018}. In this scheme, feature learning is, in fact, disconnected from the downstream optimization; the learned features thus may not be truly appropriated for the model.
He {\em et al.} further extended the deep regression idea\cite{Abhay_CNN_Surface_2018} with fully differentiable soft-argmax operations to generate surface positions followed by ReLU operations to guarantee the surface order in their fully convolutional regression network (FCRN)~\cite{YufanHe_MIA_2021}. The hybrid 2D-3D CNN \cite{LiuHong_HybridNetworkOCT_2021} using B-scan alignment was proposed to obtain continuous 3D retinal layer surfaces from OCT. The IPM optimization method \cite{Xie_IPMSurface_2022} effectively integrates the DL feature learning with the IPM optimization to enforce mutual interaction between surfaces, but the IPM optimization runs on each A-scan. 
All these methods\cite{YufanHe_MIA_2021,LiuHong_HybridNetworkOCT_2021,Xie_IPMSurface_2022} achieved highly accurate segmentation of retinal surfaces from OCT. However, their performance is prone to be affected by image outliers with bad quality or artifacts with the limited size of training data~\cite{YufanHe_MIA_2021}, as they lack the capability of explicitly learning surface smoothness structure.

This study proposes to unify the powerful feature learning capability of DL with a constrained DDP module in a single deep neural network for end-to-end training to achieve globally optimal segmentation while explicitly enforcing surface smoothness. In the proposed segmentation framework, a U-Net with additional image gradient channels~\cite{Xie_IPMSurface_2022} is leveraged as the backbone for learning parameterized surface costs. The retinal surface inference by minimizing the total surface cost while satisfying surface smoothness constraints is realized by a DDP module for a globally optimal solution. The differentiability of the DDP module enables efficient backward propagation of gradients for an end-to-end learning. To the best of our knowledge, this is the  first work to apply differentiable dynamic programming for surface segmentation in medical images. Experiments on retina spectral-domain OCT datasets demonstrated improved surface segmentation accuracy. 

\section{Method}
\vspace{-4mm}
\subsection{Problem Formulation}
\vspace{-3mm}
Let $\mathcal{I}(X, Z$ of size $X$$\times$$Z$ be a given 2D B-scan of an OCT image. For each $x = 0, 1, \ldots, X-1$, the pixel subset $\{\mathcal{I}(x, z) | 0 \le z < Z\}$ forms a column parallel to the $\mathbf{z}$-axis, denoted by $Col(x)$, which corresponds to an A-scan of the OCT image. Our goal is to seek $N  > 0$ retinal surfaces, each of which $S_i$ ($i = 0, 1,  \ldots, N-1$) intersects every column $Col(x)$ at exactly one location $z^{(i)}_x$, that is, $\mathcal{I}(x, z^{(i)}_x) \in S_i$. To find an optimal surface $S_i$, each pixel $\mathcal{I}(x, z)$ is associated with an on-surface cost $c_i(x,z)$, which is related to the likelihood of $\mathcal{I}(x,z)$ on $S_i$. Each retinal surface express a certain degree of smoothness, which specifies the maximum allowed change in the $\mathbf{z}$-dimension of a feasible surface along each unit distance change in the $\mathbf{x}$-dimension. More specifically, with given smoothness parameters $\Delta^{(i)}_x > 0$ for $S_i$, if $\mathcal{I}(x, z'), \mathcal{I}(x-1, z'') \in S_i$, then $|z' - z''| \leq \Delta^{(i)}_x$. The optimization objective of our surface segmentation problem is to maximize the total on-surface cost of all pixels on the $N$ sought surfaces $\mathcal{S}^* = \{S_0^*, S_1^*, \ldots, S_{N-1}^*\}$, with 
\vspace{-5mm}
\begin{equation} \label{eq:model}
\begin{aligned}
\mathcal{S}^* & =\operatorname*{argmax}_{\mathcal{S} = \{S_0, S_1, ..., S_{N-1}\}} \mathbb{E}(\mathcal{S})  = \sum_{i=0}^{N-1}\sum_{\mathcal{I}(x,z) \in S_i} c_i(x,z)\\
&\text{s.t.}\quad |z_x^{(i)} - z_{x+1}^{(i)} | \le \Delta_x^{(i)},  \quad \text{for  $i = 0, 1, ... , N-1$;  $x=0, 1, ... , X-2$.}\\
\end{aligned}
\end{equation}

\vspace{-8mm}
\subsection{Network Architecture}
\begin{figure}[htbp]
\vspace{-8mm}
\includegraphics[width=1.0\textwidth]{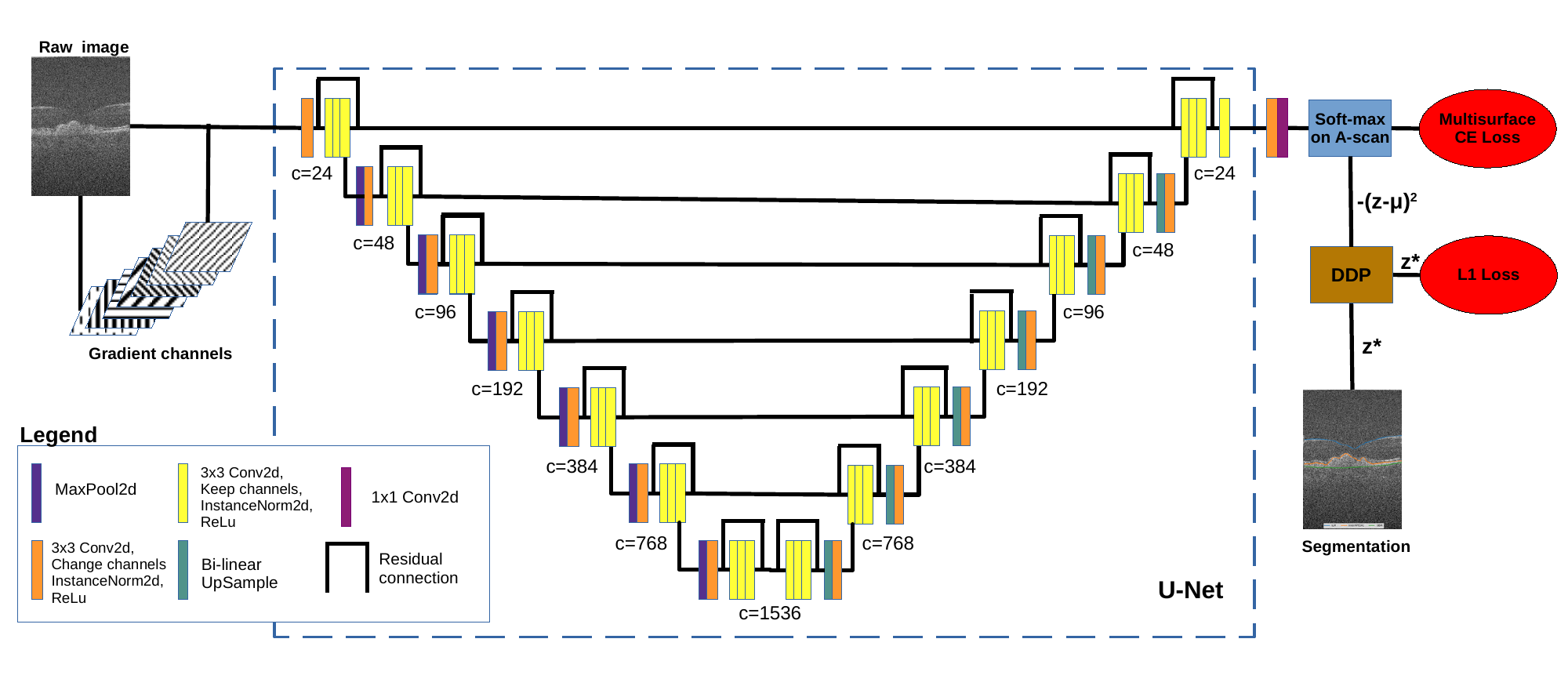}
\caption{The U-Net based network architecture with additional seven gradient channels as input. The number of channels for each layer is indicated with $c$. The DDP module solves the optimization problem for segmentation and outputs optimal smooth retinal surfaces $\mathcal{S}$ for the $L_1$ loss.}
\label{fig:networkArchitecture}
\vspace{-8mm}
\end{figure}
The proposed surface segmentation network is based on a U-Net architecture~\cite{Ronneberger_UNet_2015}, as illustrated in Fig.~\ref{fig:networkArchitecture},  which consists of seven  convolution layers. This U-Net acts as a feature-extracting module for the surface segmentation head. We started with 24 feature maps in the first convolution layer. In each downsampling layer, a conv2d module followed by a 2x2 max-pooling doubles the feature maps, and then a cascade of three same conv2d modules with a residual connection is used. The upsampling layers use a symmetric structure as the downsampling layers, but with bilinear upsample modules.  

As in the IPM segmentation method~\cite{Xie_IPMSurface_2022}, we use image gradient information as additional input channels to enrich image input information while reducing the learning burden of the network, as image gradients are prominent features to discriminate image boundaries.  Our proposed network used seven gradient channels including gradient scales along the orientations of $0^{\circ}$ ($\mathbf{x}$-dimension), $45^{\circ}$, $90^{\circ}$, and $135^{\circ}$, normalized gradient directions in the $0^{\circ}$-$90^{\circ}$ and $45^{\circ}$-$135^{\circ}$ coordinate systems, and the gradient magnitude in the $0^{\circ}$-$90^{\circ}$ coordinate system, all of which are directly computed from raw images. All these gradient channels with the raw image are then concatenated into eight channels as the input to the U-Net framework.

Followed the U-Net is a simple segmentation head, which consists of a 1x1 conv2d followed by a conv2d module. The segmentation head outputs a $N \times X \times Z$ logits. Note that $N$ is the number of target surfaces and the size of the input B-scan is $X$x$Z$. A softmax over the $\mathbf{z}$-dimension of the $N \times X \times Z$ logits gets the predicted probabilities $p^{(i)}_{x,z} \in [0,1]$, where  $i \in[0,N)$, $z \in [0,Z)$, and $x \in [0,X)$. Each $p^{(i)}_{x,z}$ indicates the probability of the pixel $\mathcal{I}(x,z)$ on Surface $S_i$. The initial estimate $\mu_x^{(i)}$ of the surface location of $S_i$ on Column $Col(x)$ can then be computed~\cite{YufanHe_MIA_2021,Xie_IPMSurface_2022}, with $\mu_x^{(i)} = \sum_{z=0}^{z=Z-1} z\cdot p^{(i)}_{x,z}$. The on-surface cost $c_i(x, z)$ for each pixel $\mathcal{I}(x,z)$ on $Col(x)$ is parameterized, as $c_i(x, z) = -(z - \mu_x^{(i)})^2$.
%

\vspace{-5mm}
\subsection{DDP Module with Smoothness Constraints}
\vspace{-1mm}
The DDP module solves the optimization problem in Eqn.~(\ref{eq:model}). The smoothness constraints $\Delta^{(i)}_x$ between two adjacent columns $Col(x)$ and $Col(x-1)$ for the target surface $S_i$ can be learned from the training data. In our experiments, $\Delta^{(i)}_x$ is simply set to be $\alpha > 0$ plus the maximum surface position difference of $S_i$ between $Col(x)$ and $Col(x+1)$ in the whole training set. Let $\tau^{(i)}_{x, z}$ denote the maximum total on-surface cost for $S_i$ starting from $Col(0)$ while ending at the pixel $\mathcal{I}(x,z)$. Based on the dynamic programming technique, we have 

%
\vspace{-3mm}
\begin{equation} 
\label{eq_bellman}
\tau^{(i)}_{x, z} =
\begin{cases}
c_i(0, z)  &\text{if x=0}\\
c_i(x, z) +\max_{z-\Delta^{(i)}_x \le z' \le z+\Delta^{(i)}_x}\{\tau^{(i)}_{x-1, z'}\} &\text{else $x \in [1,X)$},\\
\end{cases}  
\end{equation}
where $z \in [0, Z)$, and $\Delta^{(i)}_x  > 0$. The maximum of $\{\tau^{(i)}_{X-1, 0}, \tau^{(i)}_{X-1, 1}, \ldots, \tau^{(i)}_{X-1, Z-1}\}$ gives the total on-surface cost of the optimal surface $S^*_i$.

As the $\max$ operator above in Eq.~(\ref{eq_bellman}) is not differentiable,  we use a differentiable operator of LogSumExp~\cite{Mensch_DDP_2018}, as follows, to approximate the $\max$ operator:    
\begin{equation} 
\label{eq_logsumexp}
\phi^{(i)}_{x-1,z}(\tau) = \frac{1}{t} \log \sum_{z' = z-\Delta^{(i)}_x}^{z+\Delta^{(i)}_x}  \exp(t\cdot \tau^{(i)}_{x-1, z'}), \quad t > 0.
\end{equation}

This $\phi^{(i)}_{x-1,z}(\tau)$ has an elegant property that it is bounded in a narrow band of the real value $m = \max_{z-\Delta^{(i)}_x \le z' \le z+\Delta^{(i)}_x}\{\tau^{(i)}_{x-1, z'}\}$, as follows:
\begin{equation} 
\label{eq_logsumexp_bound}
m \le \phi^{(i)}_{x-1,z}(\tau) \le m + \frac{\log (2\Delta^{(i)}_x +1)}{t}
\end{equation}
In practice, we can choose a proper $t$ such that $\frac{\log (2\Delta x +1)}{t} \le \epsilon $, to make the approximation error of $\phi^{(i)}_{x-1,z}(\tau)$ is less than arbitrary small $\epsilon>0$. Using the approximation of $\phi^{(i)}_{x-1,z}(\tau)$, the DP recursive formula Eqn.~(\ref{eq_bellman}) can be written, as follows.
\begin{equation} 
\label{eq_bellman_approximation}
\tau^{(i)}_{x, z} =
\begin{cases}
c_i(0, z)  &\text{if x=0}\\
c_i(x, z) +\frac{1}{t} \log \sum_{z' = z-\Delta^{(i)}_x}^{z+\Delta^{(i)}_x}  \exp(t\cdot \tau^{(i)}_{x-1, z'} )
&\text{else $x \in [1,X)$},\\
\end{cases}  
\end{equation}
which is differentiable everywhere. The difference between our method and Mensch {\em et al.}'s DDP~\cite{Mensch_DDP_2018} is that we consider a constrained DDP model.

In the backtracking stage of DP to obtain optimal $S_i^*$, we need to know the surface location $z^{(i)}_x$ of $S_i^*$ on each column $Col(x)$. Based on Danskin's Theorem~\cite{Bertsekas_DanskinTheorem_1971}, the gradient of $\phi^{(i)}_{x-1,z}(\tau)$ attains  $\argmax_{z-\Delta^{(i)}_x\le z' \le z+\Delta^{(i)}_x}\{\tau^{(i)}_{x-1, z'}\}$, as follows:
\begin{equation} 
\label{eq_maxLocation}
\begin{aligned}
\argmax_{z-\Delta^{(i)}_x\le z' \le z+\Delta^{(i)}_x}\{\tau^{(i)}_{x-1, z'}\} &= \frac{\partial \phi^{(i)}_{x-1,z}(\tau)}{\partial \tau} &\\
& = \frac{ \exp(t\cdot \tau^{(i)}_{x-1, z'})}{\sum_{z'' = z-\Delta^{(i)}_x}^{z+\Delta^{(i)}_x}  \exp(t\cdot \tau^{(i)}_{x-1, z''})},
&\text{for $z' \in [z-\Delta^{(i)}_x, z+\Delta^{(i)}_x]$}.\\    
\end{aligned}
\end{equation}

Here Eqn.~(\ref{eq_maxLocation}) indicates that the softmax function is a smooth approximation of the argmax function. Thus, during backtracking, the optimal surface location $z^{(i)}_{x-1}$ of $S_i^*$ on $Col(x-1)$ can be computed, as follows.
\begin{equation} 
\label{eq_optimallocation_w-1}
\begin{aligned}
z^{(i)}_{x-1} = \frac{\sum_{z' = z-\Delta^{(i)}_x}^{z+\Delta^{(i)}_x}  (z' \cdot \exp(t\cdot \tau^{(i)}_{x-1, z'}))}{\sum_{z' = z-\Delta^{(i)}_x}^{z+\Delta^{(i)}_x}  \exp(t\cdot \tau^{(i)}_{x-1, z'})}.
\end{aligned}
\end{equation}

In the inference stage, we thus obtain the set of optimal surfaces $\mathcal{S}^* = \{S_0^*, S_1^*, \ldots, S_{N-1}^*\}$. While during the training stage, $\mathcal{S}^*$ is used to define the loss of the network for backward propagation.

\vspace{-4mm}
\subsection{Loss Functions}
\vspace{-2mm}
We use multiple-surface cross entropy loss $L_{mCE}$ and the L1 loss $L_1$ to train our proposed network. Let $g^{(i)}_{x,z} \in \{0,1\}$ denote the ground truth probability of pixel $\mathcal{I}(x, z)$ on $S_i$. Then, $L_{mCE}$ can be  computed, with 
\begin{equation} 
\label{eq_Loss_mce}
L_{mCE} = \frac{-\sum_{i=0}^{N-1} \sum_{z=0}^{Z-1}\sum_{x=0}^{X-1}\{g^{(i)}_{x,z}\ln(p^{(i)}_{x,z})+ (1-g^{(i)}_{x,z})\ln(1-p^{(i)}_{x,z})\}}{NZX}.  
\end{equation}

Let $s^{(i)}_{x}$ be the ground truth surface location of $S_i$ on Column $Col(x)$. The L1 loss can be computed, as follows
\begin{equation} 
\label{eq_L1}
L_{1} = \frac{1}{NX}\sum_{i=0}^{N-1}\sum_{x=0}^{X-1} \|z^{(i)}_{x} -s^{(i)}_{x} \|.  
\end{equation}

The total loss $L$ for this surface segmentation network is $L = L_{mCE} + L_{1}$. To improve the training efficiency, we first pre-train the proposed segmentation network without the DDP module, and then add it for further fine tuning of the network. In the pre-training, $z^{(i)}_{x} = \mu^{(i)}_{x}$ in Eqn.~\ref{eq_L1} is used.


\vspace{-5mm}
\section{Experiments}
\vspace{-4mm}
The proposed method was validated on two public data sets -- Duke AMD (age-related macular degeneration) dataset~\cite{Farsiu_DukeAMD_2013} and JHU MS (multiple sclerosis) dataset~\cite{YufanHe_JHUMSData_2019}.  PyTorch version 1.81 on Ubuntu Linux 20.04 was used for the implementation of the proposed method. Ablation experiments were also conducted to evaluatet the contribution of the DDP module in the proposed network. Experiments used 100 epochs for pretraining without the DDP module and then added the DDP module for further training.
\vspace{-3mm}

\subsection{Duke AMD OCT Data}
The Duke AMD data set~\cite{Farsiu_DukeAMD_2013} consists of 384 SD-OCT volumes (115 normal and 269 AMD subjects). Each original OCT volume is of size $100 \times 512 \times 1000$. The manual tracings are only available around the fovea. So each volume was cropped to form $51$ B-scans of size $512 \times 361$ around the fovea as our input data. The $\mathbf{z}$-axial resolution of A-scans is 3.24~$\mu m$/pixel. The manual tracings on each scan include three surfaces: ILM, Inter RPEDC, and OBM from top to the bottom. In our experiments, we randomly divided all 384 samples into training (187 AMD + 79 normal), validation (41 AMD + 18 normal), and test (41 AMD + 18 normal) sets.
In this experiment, we used 128 channels in the segmentation head, a batch size of 8, and an Adam optimizer with an initial learning rate of 0.1 without weight decay. 
In the fine tune stage, training with the DDP module needed 2.9 seconds per batch. 

\begin{table}[htbp]
\vspace{-5mm}
\caption{MASD ± standard deviation evaluated on Duke AMD test set and the ablation study.}
\label{table_DukeAMD_100Percent}
\centering
\resizebox{1.0\textwidth}{!}{%
\begin{tabular}{|l|c | c c c c | c c c c|}
\hline
&          &\multicolumn{4}{|c|}{Normal Group}  &\multicolumn{4}{c|}{AMD Group} \\  
Methods &TotalMean & GroupMean     &ILM    &InterRPEDC    & OBM    & GroupMean   &ILM    &InterRPEDC    & OBM \\
\hline
\multicolumn{10}{|c|}{Method comparison: MASD (mean absolute surface distance) ± standard deviation ($\mu m$)$^\dagger$}\\
\hline
G-OSC~\cite{Songqi_GraphSearch_2012}       &--         &-- & 3.85±0.16 & 4.56±0.35 &   --   & --   & 4.43±0.71 &9.33±1.74   & --\\
CNN-S~\cite{Abhay_CNN_Surface_2017}       &--         &-- & 2.88±0.22 & 4.14±0.32 &   --   & --   & 3.43±0.35 &5.92±0.84   & --\\
CNN-S-2~\cite{Abhay_CNN_Surface_2018}     &--         &4.07±0.55 & 3.36±0.23   & 3.84±0.58 &  4.97±1.01 & 5.20±1.58 & 3.71±0.77 &6.07±1.84   & 5.58±1.80\\  
FCRN~\cite{LiuHong_HybridNetworkOCT_2021}$^\ddagger$       &2.78±3.31         &-- & 1.24±0.51 & 2.06±1.51 &  2.28±0.36   & -- & 1.73±2.50 &3.09±2.09   & 4.94±5.35\\
Hybrid2D3D~\cite{LiuHong_HybridNetworkOCT_2021} &2.71±2.25         &-- & 1.26±0.47 & 2.10±1.36 &  2.40±0.39   & -- & 1.76±2.39 &3.04±1.79   & 4.43±2.68\\
Ours             &\textbf{1.88±1.96} &\textbf{1.41±0.89} &\textbf{0.55±0.41} &\textbf{1.57±0.90} &\textbf{2.10±0.45}   &\textbf{2.08±2.25} &\textbf{1.38±3.17}         &\textbf{1.80±1.10} &\textbf{3.06±1.61}\\
\hline
\multicolumn{10}{|c|}{Ablation Experiment: MASD ± standard deviation ($\mu m$)}\\
\hline
OursWithoutDDP   &1.90 ± 2.10       &1.43±0.93 & 0.56±0.24 & 1.64±1.06 &  2.10±0.46   & 2.10±2.42 & 1.47±3.65 &1.83±1.04   & 3.00±1.45\\
\hline
\multicolumn{10}{|c|}{Ablation $p\textrm{-values}$ analysis for the MASD errors}\\
\hline
                Ours vs. OursWithoutDDP     &1.43e-08   &1.06e-13      &3.23e-02 &3.82e-24 &9.78e-01           &6.81e-04     &9.23e-26 &1.00e-03   &2.24e-23\\
\hline
\multicolumn{10}{l}{$^\dagger$ Bold fonts indicate the best in its column. ``-'' indicates no reported results in the corresponding literature.}\\
\multicolumn{10}{l}{$^\ddagger$ 
The results were obtained from the re-implementation of FCRN in Ref.~\cite{LiuHong_HybridNetworkOCT_2021}.}\\
          
\end{tabular}
}

\vspace{-1mm}
\end{table}

\begin{table}[htbp]
\caption{The HD errors evaluated on Duke AMD test set. Bold fonts indicate the best in its column.}
\label{table_DukeAMD_HD}
\scriptsize 
\centering
\begin{tabular}{|l|c c c|c c c|}
\hline
                &\multicolumn{3}{|c|}{Average 95\% HD }&\multicolumn{3}{c|}{Average HD} \\
                Methods                   &ILM    &InterRPEDC    & OBM     &ILM    &InterRPEDC    & OBM   \\
                \hline
                FCRN$(\mu m)$~\cite{LiuHong_HybridNetworkOCT_2021}$^\dagger$ &4.88    &9.73          &11.32             &7.44             &16.19              &14.73  \\
                Hydrid2D3D$(\mu m)$~\cite{LiuHong_HybridNetworkOCT_2021}$^\dagger$      &4.78             &8.99          &9.91              &7.41             &14.78              &13.43  \\
                OursWithoutDDP$(\mu m)$   &4.80             &6.59          &\textbf{6.87}     &6.00             &11.13              &\textbf{9.18}  \\
                Ours$(\mu m)$             &\textbf{4.74}    &\textbf{6.53} &6.99              &\textbf{5.91}    &\textbf{10.98}     &9.25  \\
\hline
\multicolumn{7}{l}{$^\dagger$ Courtesy to Hong Liu~\cite{LiuHong_HybridNetworkOCT_2021} for the HD results. }\\

\end{tabular}
\vspace{-2mm}
\end{table}
\vspace{-6mm}

\begin{figure}[htbp]
\vspace{-6mm}
\includegraphics[width=1.0\textwidth]{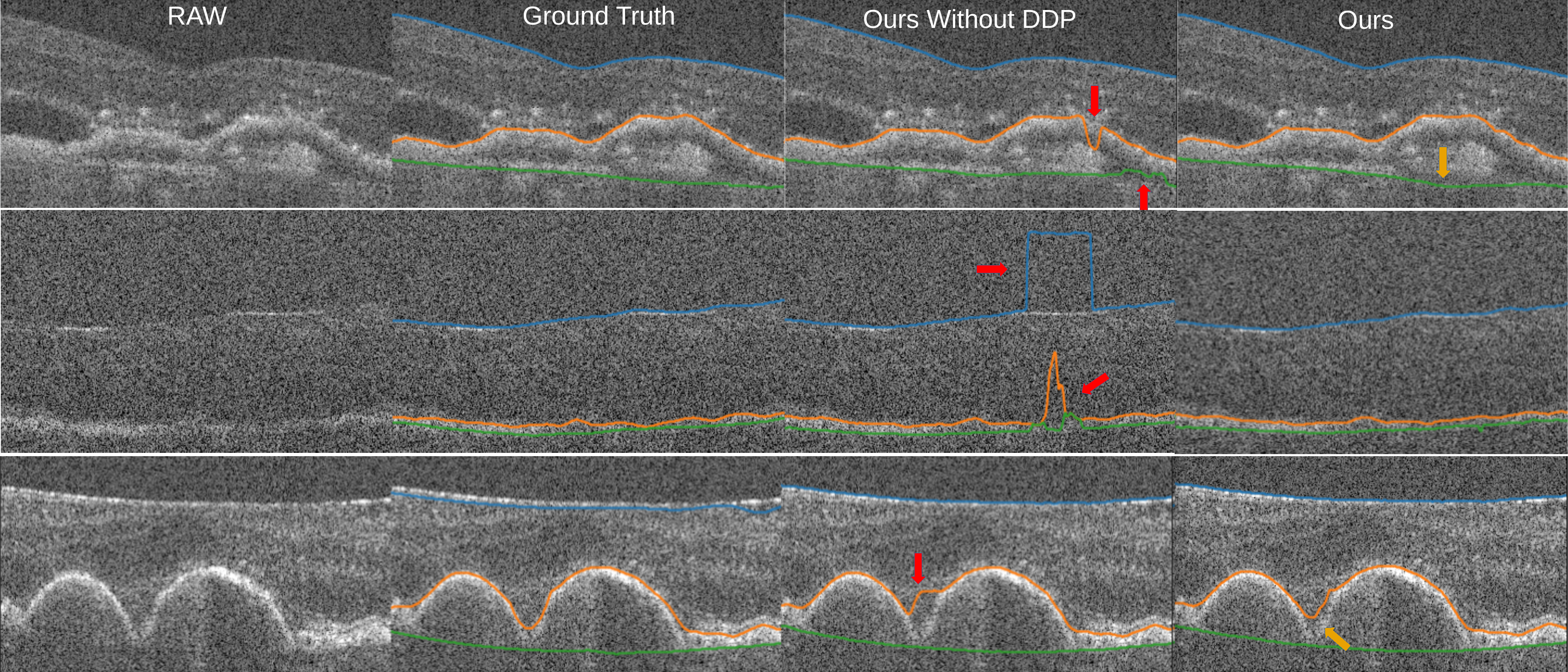}
\caption{Segmentation samples of three AMD cases in the Duke AMD test set. The red arrows indicate segmentation errors. Our proposed DDP method alleviated the errors in weak boundary regions. The orange arrows show that our DDP results are also not perfect, which show over smoothness effect on the segmented surfaces.}
\label{fig_DukeAMD_3AMDCases}
\vspace{-6mm}
\end{figure}

The mean absolute surface distance (MASD) errors for the proposed and the compared methods as well as the ablation study are shown in Table~\ref{table_DukeAMD_100Percent}. Our DDP method demonstrated improved segmentation accuracy for all MASD measurements over the compared methods. The ablation experiment and $p\textrm{-values}$ analysis showed that the DDP module significantly improved the segmentation accuracy for all three surfaces in the AMD subjects; while for the normal subjects, the DDP module significantly improved segmentation performance on both ILM and Inter RPEDC and achieved comparable results on OBM. For the Hausdorff distance (HD), our method outperformed both FCRN~\cite{YufanHe_MIA_2021} and Hybrid2D3D~\cite{LiuHong_HybridNetworkOCT_2021} (Table~\ref{table_DukeAMD_HD}), achieving much smaller HD errors especially for two challenging surfaces of Inter RPEDC and OBM.  The DDP module significantly reduced HD errors for both ILM and Inter RPEDC. Sample segmentations of three AMD cases are illustrated in Fig.~\ref{fig_DukeAMD_3AMDCases}.
\vspace{-6mm}
\subsection{JHU MS OCT Data}
The public JHU MS dataset~\cite{YufanHe_JHUMSData_2019} includes 35 human retina scans acquired on a Heidelberg Spectralis SD-OCT system, of which 14 are healthy controls (HC) and 21 have a diagnosis of multiple sclerosis (MS). Each volume has 9 surfaces, and 49 B-scans each with size of 128$\times$1024 after cropping out the center part by a Matlab script~\cite{YufanHe_JHUMSData_2019}. The $\mathbf{z}$-axial resolution of A-scans is 3.87~$\mu$m/pixel.

\begin{table}[htbp]
\vspace{-2mm}
\caption{MASD ± standard deviation evaluated on JHU MS test set and and the ablation study.}
\label{table_JHUMS_100percent}
\centering
\resizebox{1.0\textwidth}{!}{%
\begin{tabular}{|l|c|c c c c c c c c c|}
\hline
Methods                       & Overall     &    ILM       & RNFL-GCL         &    IPL-INL       & INL-OPL          & OPL-ONL          &   ELM             & IS-OS            & OS-RPE           & BM\\
\hline
\multicolumn{11}{|c|}{Method comparison: MASD (mean absolute surface distance)$^\dagger$ ± standard deviation ($\mu m$) errors}\\
\hline
FCRN~\cite{YufanHe_MIA_2021}  &2.83±1.48         &    2.41±0.81 & 2.96±1.70        &   2.87±1.69      & 3.19±1.49        &\textbf{2.72±1.70}        & 2.65±1.14         & 2.01±0.88        & 3.55±1.73        & 3.10±2.21 \\
IPM~\cite{Xie_IPMSurface_2022}&2.78±0.85       &\textbf{2.32±0.27}& 3.07±0.68        &   2.86±0.33      & 3.24±0.60        &2.73±0.57 & 2.63±0.51         &\textbf{1.97±0.57}& 3.35±0.83        &\textbf{2.88±1.63}\\
Ours                          &\textbf{2.75±0.94}&    2.35±0.38 &\textbf{2.89±0.69}&\textbf{2.83±0.41}&\textbf{3.03±0.48}& 2.74±0.62        &\textbf{2.57±0.68} & 2.02±0.80        &\textbf{3.31±0.77}& 2.99±1.99 \\
\hline
\multicolumn{11}{|c|}{Ablation Experiment: MASD ± standard deviation ($\mu m$) errors}\\
\hline
OursWithoutDDP                &    2.77±0.96   &    2.33±0.30 & 2.91±0.61        &2.79±0.39        & 3.12±0.58        &2.69±0.61         & 2.62±0.79         & 2.05±0.98        & 3.45±0.92        & 2.94±1.87 \\
\hline
\multicolumn{11}{|c|}{Ablation $p\textrm{-values}$ analysis for the MASD errors}\\
\hline
                Ours vs. OursWithoutDDP &2.30e-56  &4.94e-05 &5.20e-03 &1.59e-27 &1.47e-111 &2.33e-30 &1.95e-74 &1.13e-41 &7.21e-264 &1.65e-34  \\
\hline
\multicolumn{11}{l}{$^\dagger$ Bold fonts indicate the best in its column in method comparison.}\\
\end{tabular}
}
\vspace{-1mm}
\end{table}

\begin{table}[htbp]
\vspace{-1mm}
\caption{The ablation study on the HD errors evaluated on JHU MS test set.}
\label{table_JHUMS_HD}
\scriptsize
\centering
\resizebox{1.0\textwidth}{!}{%
\begin{tabular}{|l|c c c c c c c c c|}
\hline
Methods          &ILM &RNFL-GCL  &IPL-INL  &INL-OPL  &OPL-ONL  &ELM  &IS-OS  &OS-RPE  &BM   \\
\hline
\multicolumn{10}{|c|}{Average 95\% HD}\\
\hline
OursWithoutDDP$(\mu m)$    &5.42   &8.38      &7.24     &7.63     &7.45     &5.99  &4.66   &7.55     &5.92 \\
                Ours$(\mu m)$              &5.44   &8.11      &7.31     &7.5      &7.59     &5.85  &4.47   &7.50     &5.90 \\
                DDP's improvement &-0.37\% &3.33\%     &-0.96\%   &1.73\%    &-1.84\%   &2.39\% &4.25\%  &0.67\%    &0.34\% \\
                \hline
                \multicolumn{10}{|c|}{Average HD}\\
\hline
                OursWithoutDDP$(\mu m)$    &8.32  &13.96     &11.62    &12.03   &11.99     & 8.83  &7.85   &10.52    &8.38\\
                Ours$(\mu m)$              &8.18  &13.12     &11.33    &10.97   &11.22     & 8.25  &6.75   &10.36    &8.2 \\
                DDP's improvement &1.71\% &6.40\%     &2.56\%    &9.66\%   &6.86\%     & 7.03\% &16.30\% &1.54\%    &2.20\%\\
\hline
\end{tabular}
}
\vspace{-6mm}
\end{table}

In this experiment, the proposed segmentation model was trained on the last six HC and last nine MS subjects according to the order of their IDs and tested on the other 20 subjects, which is the same experimental configuration as in He {\em et al.}'s FCRN~\cite{YufanHe_MIA_2021}. The Gaussian, salt \& pepper noise, and random flipping on the $\mathbf{x}$-direction were used for data augmentation. The experiment used  a segmentation head of 64 channels, batch size 4, an Adam optimizer with an initial learning rate of 0.01 without weight decay, and a reducing learning rate on plateau scheduler with patience 20 and factor 0.5. In the fine tune training stage, DDP needed 13 seconds per batch. 

The MASD errors for the FCRN~\cite{YufanHe_MIA_2021}, IPM~\cite{Xie_IPMSurface_2022} , and proposed methods are shown in Table~\ref{table_JHUMS_100percent}, which also shows the ablation study results. The proposed method achieved an overall MASD error of $2.75\pm0.94$$\mu$m averaged over all nine surfaces, outperforming both compared methods. For individual surfaces, our method achieved higher segmentation accuracy on 8 surfaces compared to FCRN and on 6 surfaces compared to IPM. The DDP module significantly improved the segmentation accuracy with respect to the MASD metric for all nine surfaces. As to the HD metric, the DDP module achieved lower HD errors for all segmented surfaces and lower 95\% HD errors on six out of nine surfaces (Table~\ref{table_JHUMS_HD}). 
\vspace{-5mm}
\section{Discussion and Conclusion}
\vspace{-4mm}
In this paper, a novel DL framework for OCT surface segmentation is proposed, which unifies a constrained DDP optimization with a deep learning network for end-to-end both learning. It effectively integrates the feedback from the downstream model optimization for segmentation into the forefront feature learning by the DL network, explicitly enforcing smoothness for all segmented surfaces. 
The proposed method was validated in two public OCT datasets and outperformed the compared state-of-the-art methods. Further improvement includes using DL to learn surface smoothness constraints in the DDP module. This proposed method has the potential to be adapted for other structured surface segmentation problems in medical imaging.


\vspace{-6mm}
%
%
\bibliographystyle{splncs04}
\bibliography{reference}

\end{document}